\documentclass[a4paper,aps,prl,twocolumn,showpacs,preprintnumbers,amsmath,amssymb]{revtex4}
\usepackage{bbm}
\usepackage{amsmath}
\usepackage{verbatim}
\usepackage{graphicx}
\bibliography{References}
\begin{document}

\title{Quantum processing photonic states in optical lattices}

\author{Christine A. Muschik, In\'es de Vega, Diego Porras, and J. Ignacio Cirac$^1$}

\affiliation{ $^1$Max-Planck--Institut f\"ur Quantenoptik,
Hans-Kopfermann-Strasse, D-85748 Garching, Germany}

\begin{abstract}
The mapping of photonic states to collective excitations of atomic
ensembles is a powerful tool which finds a useful application in the
realization of quantum memories and quantum repeaters. In this work
we show that cold atoms in optical lattices can be used to perform
an entangling unitary operation on the transferred atomic
excitations. After the release of the quantum atomic state, our
protocol results in a deterministic two qubit gate for photons. The
proposed scheme is feasible with current experimental techniques and
robust against the dominant sources of noise.
\end{abstract}

\pacs{42.50.-p, 03.67.Lx, 03.67.Mn, 32.80.Qk}


 \maketitle

%
%

Photonic channels are ideally suited for the transmission of quantum
states, since current technology is able to
distribute photons between remote locations by means of optical
fibers. For this reason, they play a key role in practical
applications of quantum information such as quantum cryptography.
The storage and manipulation of photons is, however, more
problematic.
Storage of photonic quantum states can be efficiently implemented by
interfacing a photonic channel with an atomic system. This idea can
be used to realize quantum repeaters \cite{QuR1}, and thus, to
overcome the problem of losses in the photonic channel by applying
entanglement purification at intermediate locations.
Manipulation of photonic states requires the ability to perform
entangling operations. One possibility is to make use of materials
which exhibit optical nonlinearites, but so far, available
nonlinearities are too weak to provide us with short gate times.
A completely  different approach which only requires linear optical
operations and measurements was proposed in \cite{KLM}. However,
this scheme is not very efficient in practice.

In this work we show how to perform a deterministic entangling gate
between photons by interphasing a system of cold neutral atoms in an
optical lattice with a photonic channel; that is, we show that this
atomic system can perform at the same time the tasks of storing and
processing quantum information.
The atomic ensemble is assumed to be in a Mott insulating phase such
that the lattice is filled with approximately one atom per site
\cite{MI}.
The photonic input state is mapped to a collective atomic state
following light--matter interface schemes \cite{QM}. The ability to
control atomic interactions in the optical lattice allows us to
perform a gate on collective atomic states which are then released
back to the photonic channel.
In this way, our proposal profits from the  advantages of two
different experimental techniques which have been recently
demonstrated.

We consider qubits defined by the absence or the presence of a
photon, but with a few modifications our scheme is also well suited
to process polarization qubits.
Our gate transforms the light input state
$|\Psi^{\text{in}}\rangle_{\text{L}}=\alpha
|0\rangle_{\text{L}}+\beta|1\rangle_{\text{L}} +\gamma
|2\rangle_{\text{L}}$ into
$|\Psi^{\text{out}}\rangle_{\text{L}}=\alpha
|0\rangle_{\text{L}}+i\beta|1\rangle_{\text{L}} +\gamma
|2\rangle_{\text{L}}$, where $|n\rangle_{\text{L}}$ is the $n$
photon Fock state.
Together with one-qubit rotations this operation is sufficient for
universal quantum computation \cite{UnivSet}.
Each incoming photon creates a collective atomic state, within the
subspace of excitations coupling to the light state. This state has
to be manipulated in such a way that the resulting state belongs to
the same subspace. Using controlled collisions between atoms, this
task would require a large number of  $O(N^2)$  operations, since
each atom has to interact with all the others. We face here the
problem of implementing efficiently a nonlinear operation with
collective states, having only local interactions at our disposal.
With our scheme we manage to reduce the number of operations to
$O(N^{1/3})$, by reducing the effective dimensionality of the
problem. The main idea is to map collective excitations from the
three dimensional Mott insulator to a plane of particles, then to a
line and finally to a single atom, which can be directly
manipulated. The plane, the line and the single atom are created by
means of an initialization protocol, which has to be performed once,
before quantum gates can be run on the lattice.
Remarkably, our proposal does not require addressability of
individual atoms, and involves only two internal atomic levels. It
comprises four kinds of basic operations, which are all within the
experimental state of the art. Finally, the scheme is robust against
the main sources of errors in a realistic setup.
%
%
%
%
%
%

Atoms are assumed to possess two internal states $|a\rangle$ and
$|b\rangle$ and to be initially prepared in $|a\rangle$.
As in quantum memory protocols, for example in \cite{QM}, the
photonic input state is transferred to the atomic ensemble such that
photonic Fock states $|n\rangle_{\text{L}}$ are mapped to collective
atomic states with $n$ excitations $|n\rangle_{\text{A}}$. The
initial atomic state is therefore given by
$|\Psi^{\text{in}}\rangle_{\text{A}}=\alpha |0\rangle_{\text{A}}
+\beta |1\rangle_{\text{A}}+\gamma |2\rangle_{\text{A}}$, where
$|1\rangle_{\text{A}}$ is a superposition of all permutations of $N$
particle product states containing one atom in $|b\rangle$,
$|1\rangle_{\text{A}}=\sum_{j=1}^{N}f_j|a\rangle_{1}...|b\rangle_{j}...|a\rangle_{N}$
with $\sum_j^N |f_j|^{2}=1$. An analogous definition holds for
$|2\rangle_\text{A}$.
%
%
%

Atomic states are processed by means of the following four
operations.
(1) {\it State--dependent transport.}
Atoms are displaced depending on their internal state using optical
lattices with different polarizations \cite{BCJD99,
JBCGZ99+MGWRHB03}.
(2) {\it Population transfer between atomic states.}
Coherent coupling of the two atomic levels is achieved by driving
Rabi oscillations. A $\pi/2$ pulse creates the coherent
superposition
$|a\rangle\mapsto(|b\rangle-|a\rangle)/\sqrt{2}$,
$|b\rangle\mapsto(|b\rangle+|a\rangle)/\sqrt{2}$,
while a $\pi$ pulse inverts the atomic population.
(3) {\it Collisional phase shift.}
Controlled collisions between particles in different states are
induced by spin dependent transport. If two particles occupy the
same lattice site a collisional phase $\phi_{\textmd{col}}= \Delta E
\ t_{\textmd{int}}$ is accumulated \cite{JBCGZ99+MGWRHB03}, where
$\Delta E$ is the on--site interaction. By controlling the
interaction time $t_{\textmd{int}}$, $\phi_{\textmd{col}}$ can be
tuned.
(4) {\it State-dependent phase shift.}
A state-dependent single particle rotation can be obtained, for
example, by applying a magnetic field.
%
%
%

By combining these elements, the two qubit CNOT operation can be
implemented (see \cite{JBCGZ99+MGWRHB03} for details). It transfers
a target atom from its initial state $|a\rangle_\textmd{t}$ to
$|b\rangle_\textmd{t}$ if the control atom is in
$|b\rangle_\textmd{c}$. More specifically, consider control and
target atoms placed along the $x$--axis at $x_\textmd{c}$ and
$x_\textmd{t}$ $( > x_\textmd{c} )$, respectively. First, a $\pi/2$
pulse is applied to the target atom $| a \rangle_\textmd{t}
\rightarrow (|b\rangle_\textmd{t} - |a\rangle_\textmd{t})/\sqrt{2}$.
Then the $| b \rangle$ lattice is displaced along $\hat{x}$, such
that the control atom collides with the target atom and induces a
$\pi$ phase on $| a \rangle_\textmd{t}$. Finally, the initial
positions of the atoms are restored and a second $\pi/2$ pulse is
applied to the target atom $(|b \rangle_\textmd{t} +
|a\rangle_{\textmd{t}})/\sqrt{2} \rightarrow | b
\rangle_{\textmd{t}}$.
%
%
%
%

\begin{figure}[pbt]
\begin{center}
\includegraphics[width=8.8cm]{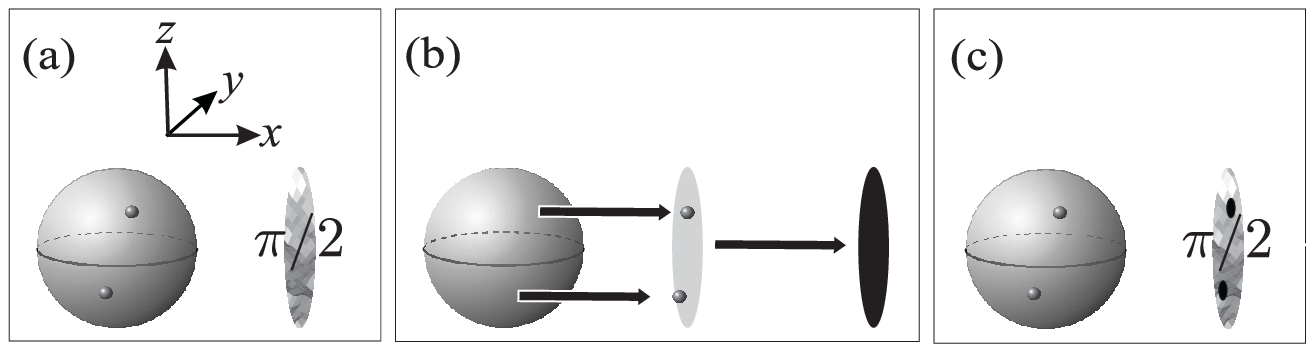}
\caption{Mapping of excitations in the bulk to the plane. (a) A
$\pi/2$ pulse is applied to the plane. (b) The $|b\rangle$ lattice
is shifted along $\hat{x}$ such that atoms in $|b\rangle$ in the
bulk interact with the $|a\rangle$ part of the plane. The time spent
after each single site displacement is chosen such that a phase
$\pi/2$ is accumulated if a collision occurs. Then this lattice
movement is reversed. Thus the initial positions of the atoms are
restored and each target atom which is located on the path of a
control atom in $|b\rangle$ experienced two collisions and picked up
a total phase of $\pi$. (c) Finally the plane is subjected to
another $\pi/2$ pulse, which transfers most of the atoms back to
$|a\rangle$. Only atoms, which suffered a collision are transferred
to $|b\rangle$.} \label{Mapping}
\end{center}
\end{figure}

\begin{figure}[pbt]
\begin{center}
\includegraphics[width=8.8cm]{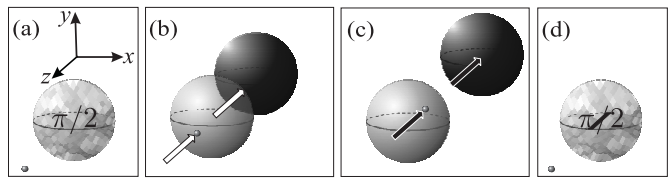}
\caption{Creation of a line of atoms in $|b\rangle$. (a) A $\pi/2$
pulse is applied to the target atoms. (b) $|a\rangle$ and
$|b\rangle$ components of the target qubits are separated spatially
by a $|b\rangle$ lattice shift along  $-\hat{z}$. (c) The
$|b\rangle$ lattice is further displaced along $-\hat{z}$, such that
the control atom in $|b\rangle$ interacts successively with the
$|a\rangle$ part of target atoms along its path, each time leading
to a collisional phase $\pi/2$. Both lattice shifts are reversed
leaving all atoms in their original positions. (d) A $\pi/2$ pulse
is applied to the target qubits. Atoms which have interacted with
the control atom are transferred to $|b\rangle$.}
\label{Columnization}
\end{center}
\end{figure}

The key idea in our scheme is to move control atoms in $| b \rangle$
through a set of target atoms in $|a\rangle$, thus transferring the
atoms along its path to state $| b \rangle$. This tool is employed
in two related procedures, which lie at the heart of the proposed
scheme and are introduced now.
(I) {\it Mapping of collective excitations from an atomic ensemble
of dimensionality $d$ to a sample of dimensionality $d-1$.}
A set of control qubits acts upon a set of target qubits. An example
is illustrated in Fig. \ref{Mapping}, where control atoms in a three
dimensional Mott insulator act on target atoms in a plane.
If an atom in the bulk is in state $|b\rangle$, a collision is
induced and the target atom hit by the control atom along its path
through the plane is transformed to $|b\rangle$. In this way atoms
in $|b\rangle$ are projected from the bulk to the plane.
More precisely, the procedure maps a state with $n$ atoms in
$|b\rangle$ to a state with $n$ atoms in $|b\rangle$, except if two
atoms in $|b\rangle$ in the bulk are located in a line along
$\hat{x}$, leaving the corresponding target atom in $|a\rangle$
($CNOT^{2}=1$). In any case an even/odd number of excitations is
mapped to an even/odd number of excitations in the target object.
This method allows us to reduce stepwise the dimensionality of the
problem. In the last step excitations are mapped from a line to a
single site ($d=1$), and an odd number of excitations in the line
transfers the target atom to state $|b\rangle$, while in case of an
even number of excitations this atom is left in state $|a\rangle$.
Thus the parity information is encoded in the state of a single
atom.
(II) {\it Creation of a $d$ dimensional structure from a $d-1$
dimensional one.}
Starting from a control atom in $|b\rangle$ and an ensemble of
target atoms in $|a\rangle$, a line of atoms in $|b\rangle$ can be
produced by running many CNOT operations in series, such that the
control qubit in $|b\rangle$ acts successively on several target
atoms in a row, which are accordingly transferred to state
$|b\rangle$ as explained in Fig. \ref{Columnization}. \footnote{The
separation step shown in \ref{Columnization}b can be performed
either by displacing the lattices first by half a lattice spacing
along $\hat{x}$ (or $\hat{y}$) and then by a distance exceeding the
length of the ensemble along $-\hat{z}$, or by moving the lattice
fast along $-\hat{z}$, which can be done such the atoms start and
end up in their motional ground state \cite{BCJCZ99}.}
For the purpose of producing a plane (d=2), we proceed analogously
with a line of atoms in state $|b\rangle$ instead of a single
control qubit.
%
%
%
%
%
%
%
%
%
%

\begin{figure}[pbt]
\begin{center}
\includegraphics[width=8.8cm]{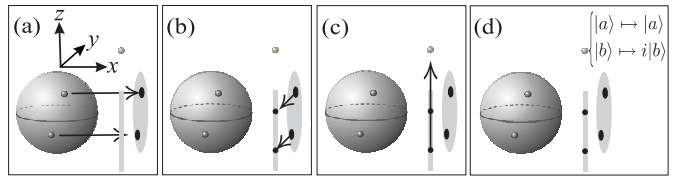}
\caption{Quantum gate protocol transforming the input state
$|\Psi^{\text{in}}\rangle_{\text{A}}= \alpha
|0\rangle_{\text{A}}+\beta|1\rangle_{\text{A}}+\gamma|2\rangle_{\text{A}}$
into $|\Psi^{\text{out}}\rangle_{\text{A}}=\alpha
|0\rangle_{\text{A}}+i\beta|1\rangle_{\text{A}}+\gamma|2\rangle_{\text{A}}$.
(a)-(c) Excitations in the Mott insulator are successively mapped to
structures of lower dimensionality resulting in a single atom being
in state $|a\rangle$/$|b\rangle$ in case of an even/odd number of
excitations in the Mott insulator. (d) A state dependent phase is
applied to the isolated particle such that
$|1\rangle_{\text{A}}\mapsto i|1\rangle_{\text{A}}$. Subsequently
steps (a)-(c) have to be reversed.}\label{Protocol}
\end{center}
\end{figure}

As mentioned above, the whole scheme can be decomposed into two
phases. During the initialization phase, the atoms are divided into
four sets, namely the bulk, a plane, a line, and a dot, which are
spatially separated. This setup has to be prepared once and can
afterwards be used many times to perform gates. In the second phase
the quantum gate protocol itself is performed. We explain first the
processing part and then how the setup is prepared.
%
%
%
The quantum gate protocol is summarized in Fig. \ref{Protocol}.
The parity of the number of excitations contained in the bulk is
mapped to the dot by means of procedure (I), such that the isolated
atom is in state $|b\rangle$ in case of one excitation, while it is
in state $|a\rangle$ otherwise. Now a phase shift $\pi/2$ is applied
to the dot if it is in state $|b\rangle$. In this way, the atomic
state is transformed according to $
|0\rangle_{\text{A}}\mapsto|0\rangle_{\text{A}}$,
$|1\rangle_{\text{A}}\mapsto i|1\rangle_{\text{A}}$,
$|2\rangle_{\text{A}}\mapsto|2\rangle_{\text{A}}$.
Then, the previous steps are reversed and the excitations are
converted to light, leaving the setup in the original state. Note
that none of these steps requires addressability of single sites.
%
%
Now we consider the truth table corresponding to the protocol. Let
us denote by $|B^{n}_k\rangle_\text{b}$ the initial state of the
bulk containing  $n=0,1,2$ atoms in $|b\rangle$, located at certain
lattice sites according to a configuration $k$ and by
$|P\rangle_\text{p}$, $|L\rangle_\text{l}$ the state of the plane
and the line with all atoms in state $|a\rangle$. Procedure (I)
produces the map
\begin{eqnarray*}
|B^{n}_k\rangle_\text{b}|P\rangle_\text{p}|L\rangle_\text{l}|a\rangle_\text{d}
\mapsto
|B^{n}_k\rangle_\text{b}|P^{n'}_k\rangle_\text{p}|L^{n''}_k\rangle_\text{l}|a_n\rangle_\text{d},
\end{eqnarray*}
where $|P^{n'}_k\rangle_\text{p}$ and $|L^{n''}_k\rangle_\text{l}$
refer to the states of the plane and line after the excitations have
been mapped and $|a_n\rangle_\text{d}$ describes the state of the
dot with $a_0=a_2=a$ and $a_1=b$. Thus, the whole protocol results
in
 \begin{eqnarray}
 |B^{n}_k\rangle_b|P\rangle_p|L\rangle_l|a\rangle_\text{d} \mapsto
 i^{n {\rm mod
 2}}|B^{n}_k\rangle_b|P\rangle_p|L\rangle_l|a\rangle_\text{d}.\
 \label{BB}
 \end{eqnarray}
%
%
%

\begin{figure}[pbt]
\begin{center}
\includegraphics[width=8.8cm]{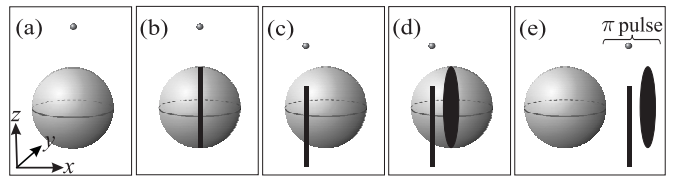}
\caption{Initialization of the lattice. (a) A control atom in
$|b\rangle$ is placed outside the ensemble. (b) The control qubit
interacts successively with a row of target atoms in the ensemble,
thus transferring them to state $|b\rangle$, as explained in figure
\ref{Columnization}. We obtain a line of atoms in $|b\rangle$, which
is aligned along $\hat{z}$. (c) The line is separated from the
ensemble along $\hat{-y}$ . (d) The line of control qubits is now
used to create a plane of atoms in $|b\rangle$. For this purpose a
$\pi/2$ pulse is applied to the ensemble, collisions are induced by
a $|b\rangle$ lattice shift along $\hat{y}$ and another $\pi/2$
pulse is applied to the bulk. Since each control atom in the line
leads to a line of atoms in $|b\rangle$, which is aligned along
$\hat{y}$ we obtain a plane in the $\hat{x}\hat{y}$ plane. (e) The
plane is separated from the ensemble by a $|b\rangle$ lattice shift
along $\hat{x}$.}\label{Initialization}
\end{center}
\end{figure}

The initialization protocol is summed up in Fig.
\ref{Initialization}. First a collective excitation
$|1\rangle_\textmd{A}$ is created\footnote{This can for example be
done using heralded single photons from an EPR source or a weak
coherent field together with a postselecting photon detection.}.
This state contains one atom in $|b\rangle$, which is separated from
the ensemble and subsequently used to create a line of atoms in
$|b\rangle$ following procedure (II). Next, this line is separated
from the bulk and utilized to produce a plane of atoms in
$|b\rangle$ employing the same method. Finally the plane is
displaced such that the constellation shown in Fig.
\ref{Initialization}e is obtained, and a $\pi$ pulse is applied to
the plane, the line and the dot transferring these atoms to state
$|a\rangle$.

Note that collective excitations are not localized, but a
superposition of states with atoms in $|b\rangle$ at different
sites. Moreover, we have a superposition of different positions of
the plane, the line and the dot, as the excitation created at the
beginning of the initialization is also delocalized.
For any term in the superposition, the final state differs only in a
phase from the initial state. By adding the terms in equation
(\ref{BB}) with respect to the positions of the excitations, $k$,
and the positions of the plane, the line and the dot we obtain the
desired quantum gate transformation.
%
%

In the following we analyze the main sources of errors in our scheme
\footnote{A quantitative analysis will be given elsewhere.}. It has
been carefully designed in order to minimize decoherence, first of
all, by avoiding the presence of cat states in the internal atomic
states, which would give rise to errors if few particles are lost.
Apart from that, runtimes are very short such that decoherence has
not much time to act.
In particular, the time required to perform the scheme is
essentially given by the time needed to run the collisional steps,
since population transfers and separations can be done much faster.
Each collisonal step has to be performed along a whole ensemble
length and requires therefore a time $t_{int} N^{1/3}$, where
$t_{int}$ is the time spent in a single collision. Remarkably, the
three dimensional problem scales like a one dimensional one in time,
since the task of scanning $N$ particles in a three dimensional
lattice is accomplished by a one dimensional projection scheme.
%
%
%
%
%
%
%
%
%
%
%

We address now transitions from $|a\rangle$ to $|b\rangle$ or to
another trapped state affected by $|b\rangle$ lattice shifts and
give an example how a judicious choice of atomic levels allows us to
sidestep this source of errors, while still being able to perform
the operations that are necessary for the quantum gate.
%
%
By employing alkali atoms with nuclear spin $1/2$ for instance, the
atomic qubit can be encoded in hyperfine states of the $S_{1/2}$
shell by identifying $|a\rangle \equiv |F=1, m_{F}=-1\rangle$, and
$|b\rangle \equiv |F=1,m_{F}=1\rangle$.
By choosing appropriately the detuning of two off--resonant standing
waves with different polarizations \cite{JBCGZ99+MGWRHB03},
state--dependent transport can be implemented by trapping
$|F=1,m_{F}=-1\rangle$ and $|F=1, m_{F}=1\rangle$ by $\sigma_{-}$
and $\sigma_{+}$ light respectively.
Transitions $|a\rangle\mapsto |b\rangle$ cannot be induced by means
of the off--resonant laser fields, since $|a\rangle$ corresponds to
the nuclear magnetic quantum number $m_{I}=-1/2$, while $|b\rangle$
corresponds to $m_{I}=1/2$. $\pi/2$ or $\pi$ pulses can be applied
by means of resonant two-photon Raman or microwave transitions.
Finally, the standing waves do induce transitions to the other
trapped states  $|F=1,m_{F}=0\rangle$ and $|F=0,m_{F}=0\rangle$.
However, the optical potential experienced by these levels is given
by the equally weighted sum of contributions from both
polarizations. While shifting one lattice with respect to the other,
the optical potential vanishes at some point, and these two levels
are emptied, which ensures that they do not affect the protocol.

%
%
%
Among the remaining noise mechanisms, the most important ones are
imperfect population transfer and dephasing of quantum states due to
spontaneous emission\footnote{Inhomogeneous background fields lead
to uncontrolled relative phases
$(|a\rangle+e^{i\beta}|b\rangle)/\sqrt{2}$ during the state
dependent transport, thus inducing dephasing. This effect can be
suppressed, since all transport shifts are reversed in each mapping
step. By swapping $|a\rangle$ and $|b\rangle$ by means of a $\pi$
pulse between the first and the second (reversing) shift both parts
of the superposition $|a\rangle$ and $|b\rangle$ acquire the same
phase.} between two $\pi/2$ pulses.
The corresponding probability of error is proportional to the number
of target atoms in the mapping steps $N^{2/3}$.
This failure probability can be reduced by using an elongated atomic
ensemble having a spatial extend $L$ along the direction of the
first lattice shift in the quantum gate protocol and a length $l<L$
along the other directions. In this case the probability of
obtaining a wrong result is proportional to $l^{2}$.
%
%
%

The probability of error due to the remaining noise mechanism scales
at worst like $N^{1/3}$, i.e. proportional to the runtime of the
protocol. First we consider imperfections in the $\pi$ pulse, which
is performed at the end of the initialization of the lattice. Since
an imperfect population transfer leaves atoms in a superposition
state, the $| b\rangle$ lattice should be emptied as an additional
step of the initialization after the $\pi$ pulse.
%
%
Another source of errors are occupation number defects. We only have
to deal with empty lattice sites, since double occupied sites can be
avoided by choosing low filling factors. Holes in the plane and the
line lead to a wrong result, if they are located at specific sites
which interact with an atom in $|b\rangle$ in the course of the
processing protocol. The failure probability due to defects which
are initially present in the Mott insulator are given by the
probability for a single site to be unoccupied, and does not depend
on the size of the system.
Holes can also be created as consequence of atomic transitions into
untrapped states. This dynamical particle loss induces an error
which scales like the duration of the gate, $N^{1/3}$.
%
%
Another limiting factor are imperfect collisions. The phase acquired
in each lattice shift during the collisional steps may differ from
$\phi_{\textmd{col}}=\pi$. However, as in the case of unoccupied
lattice sites, the probability of obtaining a wrong result due to
such an event is given by the probability on the single-site level.
%
%
%
The fidelity of the scheme is also decreased by undesired
collisional phases.
The corresponding failure probability is proportional to $N^{1/3}$,
since these phases are accumulated in one dimensional operations
each covering one ensemble length.
%
%
Finally, kinetic phases acquired by the atoms during lattice shifts
do not play a role in the proposed scheme. Employing the common
technique for state dependent transport, the nodes of two optical
potentials forming standing waves are moved in opposite directions
$V_{\pm}(x)=\cos^{2}(kx\pm\phi)$ for some wave vector $k$, spatial
variable $x$ and angle $\phi$. Lattice shifts affect therefore both
atomic species in the same way and lead only to global phases of the
resulting quantum states.

We thank Eugene Polzik for discussions and acknowledge support from
the Elite Network of Bavaria (ENB) project QCCC, the EU projects
SCALA and COVAQUIAL, the DFG-Forschungsgruppe 635 and Ministerio de
Educacion y Ciencia EX-2006-0295.
%
%

%
\end{document}